\documentclass{aastex62}

\usepackage{natbib}

\graphicspath{{./}{figures/}}

\accepted{Astronomical Journal}

\shorttitle{Multiple Transits during a Single Conjunction}
\shortauthors{Kostov et al.}

\newcommand{\tess}{{TESS}}

\newcommand{\kepler}{{\it Kepler~}}

\newcommand{\vcalcx}{$(V_{\rm calc})_x$}
\newcommand{\pcalc}{${P_{\rm calc}}$}
\newcommand{\ptrue}{${P_{\rm true}}$}

\begin{document}

\title{Multiple Transits during a Single Conjunction: Identifying Transiting Circumbinary Planetary Candidates from \tess}

\correspondingauthor{Veselin Kostov}
\email{veselin.b.kostov@nasa.gov}

\author[0000-0001-9786-1031]{Veselin~B.~Kostov}
\affiliation{NASA Goddard Space Flight Center, 8800 Greenbelt Road, Greenbelt, MD 20771, USA}
\affiliation{SETI Institute, 189 Bernardo Ave, Suite 200, Mountain View, CA 94043, USA}
\author[0000-0003-2381-5301]{William F. Welsh}
\affil{Department of Astronomy, San Diego State University, 5500 Campanile Drive, San Diego, CA 92182, USA}
\author{Nader Haghighipour}
\affil{Institute for Astronomy, University of Hawaii-Manoa, Honolulu, HI 96822, USA}
\author[0000-0002-9644-8330]{Billy Quarles}
\affil{Center for Relativistic Astrophysics, School of Physics, 
Georgia Institute of Technology, 
Atlanta, GA 30332, USA}
\author{Eric Agol}
\affil{Department of Astronomy, University of Washington, Seattle, WA 98195, USA}
\author{Laurance Doyle}
\affiliation{SETI Institute, 189 Bernardo Ave, Suite 200, Mountain View, CA 94043, USA}
\author{Daniel C. Fabrycky}
\affil{Department of Astronomy and Astrophysics, University of
Chicago, 5640 S. Ellis Ave, Chicago, IL 60637, USA}
\author{Gongjie Li}
\affil{Center for Relativistic Astrophysics, School of Physics, 
Georgia Institute of Technology, 
Atlanta, GA 30332, USA}
\author[0000-0002-7595-6360]{David~V.~Martin}
\altaffiliation{Fellow of the Swiss National Science Foundation}
\affil{Department of Astronomy and Astrophysics, University of
Chicago, 5640 S. Ellis Ave, Chicago, IL 60637, USA}
\author{Sean Mills}
\affil{Department of Astronomy, California Institute of Technology, Pasadena CA, 91125, USA}
\author{Tsevi Mazeh}
\affil{Department of Astronomy and Astrophysics, Tel Aviv University, 69978 Tel Aviv, Israel}
\author[0000-0001-9647-2886]{Jerome A. Orosz}
\affil{Department of Astronomy, San Diego State University, 5500 Campanile Drive, San Diego, CA 92182, USA}
\author{Brian P. Powell}
\affiliation{NASA Goddard Space Flight Center, 8800 Greenbelt Road, Greenbelt, MD 20771, USA}

\begin{abstract}

We present results of a study on identifying circumbinary planet candidates that produce multiple transits during one conjunction with eclipsing binary systems. The occurrence of these transits enables us to estimate the candidates' orbital periods, which is crucial as the periods of the currently known transiting circumbinary planets are significantly longer than the typical observational baseline of \tess. Combined with the derived radii, it also provides valuable information needed for follow-up observations and subsequent confirmation of a large number of circumbinary planet candidates from \tess. Motivated by the discovery of the 1108-day circumbinary planet Kepler-1647, we show the application of this technique to four of {\em Kepler's} circumbinary planets that produce such transits. Our results indicate that in systems where the circumbinary planet is on a low-eccentricity orbit, the estimated planetary orbital period is within $<10-20\%$ of the true value. This estimate is derived from photometric observations spanning less than 5\% of the planet's period, demonstrating the strong capability of the technique. Capitalizing on the current and future eclipsing binaries monitored by NASA's \tess\ mission, we estimate that hundreds of circumbinary planets candidates producing multiple transits during one conjunction will be detected in the \tess\ data. Such a large sample will enable statistical understanding of the population of planets orbiting binary stars and shed new light on their formation and evolution. 

\end{abstract}


\section{Introduction\label{sec:intro}}

NASA's {\it Kepler} and \tess\ missions have ushered into an exciting era of exoplanetary science by enabling, for the first time, the detection of planets transiting main-sequence binaries. Known as Circumbinary Planets (CBPs), these detections strongly indicate that planet formation around binary systems is robust, and that planets of a variety of sizes and orbital configurations may exist in such dynamically complex environments. 

Today, we know of 13 transiting CBPs in 11 different systems, all discovered around eclipsing binary stars that have periods longer than 7.5 days (Welsh \& Orosz 2018, Kostov et al. 2020). It is widely accepted that these planets formed at distances beyond their current orbits and migrated to their present locations (e.g.~Kley \& Haghighipour 2014). Those that stopped their migration in orbits between mean-motion resonances with their host binaries managed to avoid the destructive nature of these resonances, and maintained long-term stable orbits. Many others might have been scattered out or crashed into the central binary (e.g.,~Sutherland \& Fabrycky 2016).

The current population of CBPs, although small, has shown some interesting characteristics. For instance, 9 of the 10 \kepler CBP systems have planets that orbit within a factor of two of the location of the boundary of orbital instability around their host binaries. The orbits of all 13 currently known transiting CBPs are within a few degrees of the planes of their corresponding binaries (although there may be a strong selection bias) and precess on timescales ranging from decades (Kepler-413 b, precession period = 11 years) to millennia (Kepler-1647 b, precession period $>$ 7000 years). Compared to transiting planets around single stars, the transiting CBPs have on average longer orbital periods (the longest known transiting system is the CBP Kepler-1647 b, with an orbital period of $\sim 1100$ days).  All of the currently known transiting CBPs have radii between that of Neptune and Jupiter (a size-range for which relatively few single-star planets exist), and 4 are in the habitable zone (Haghighipour \& Kaltenegger 2013, Welsh \& Orosz 2018, Martin 2019).

It is important to emphasize that because the number of transiting CBPs is small, generalizing their characteristics to all such planets is premature. In order to be able to make such a generalization, more CBPs need to be discovered so that their orbital and physical properties can be studied statistically. Fortunately, theoretical models point to a high efficiency for planet formation in circumbinary disks (e.g.~Kley \& Haghighipour 2014). However, the combination of planet migration and planet-planet scattering may place these planets in non-transiting configurations, reducing the efficiency of their detection using transit photometry (e.g.~Pierens \& Nelson 2013; Bromley \& Kenyon 2015; Kley \& Haghighipour 2015). Simple geometrical arguments show that, for instance, for each of the transiting CBPs discovered to date, there must be many more that did not transit during the time of the observation because of unfavorable orbital configuration. In addition, even if their orientations were near edge-on, many of the transiting systems were not at the appropriate phase of their precession cycle to exhibit transits during their observation (e.g.~Schneider 1994, Welsh et al. 2012, 2015; Kostov et al. 2014, 2016, 2020; Martin 2017). 

With its all-sky coverage, NASA's \tess\ space telescope presents a promising pathway to the detection of many more transiting CBPs. However, because \tess\ observations are ${\rm \sim28\ days}$ in duration, it will not be possible to follow the same discovery process as the one used for \kepler CBPs (i.e.,~detecting transits from several orbital cycles of the planet). If \tess\ observations are to be used for detecting CBPs, their transits must occur within the 28-day window --- in other words, multiple transits during a single conjunction. 

\begin{figure*}
\includegraphics[scale=0.15]{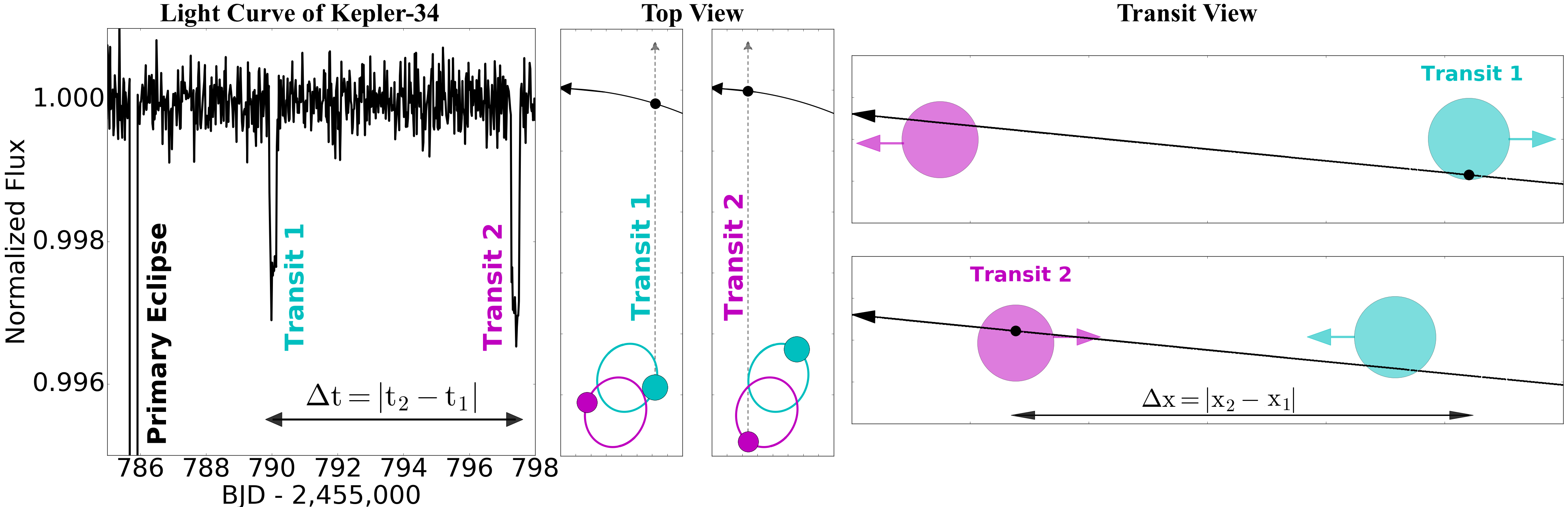}
\caption{{\it Left}: A short segment of the {\it Kepler} light curve of Kepler-34 showing two transits of the CBP during one conjunction. 
{\it Middle and Right}: Schematic orbital configuration of the system during the two transits.
The dashed arrow in the middle panels indicates the direction to the observer. The right panels show 
the orbital configuration of the system as seen from the {\it Kepler} telescope. 
The size of the CBP (black circle) has been exaggerated.
\label{fig:k34_LC}}
\end{figure*}

Interestingly, such multiple-transit, single-conjunction events have already been detected in the light-curves of 4 out of 11 \kepler CBP systems, Kepler-16, 34, 35, and 1647, demonstrating that this is not a rare occurrence. Figure \ref{fig:k34_LC} shows this for the Kepler-34 system, where the CBP transits each star of its host binary within the span of a few days. As discussed by Schneider \& Chevreton (1990) and demonstrated by Kostov et al. (2016), the occurrence of such transits can be used to identify a CBP candidate and estimate its orbital period. This motivated us to further investigate this method as a potential mechanism for detecting CBP candidates. In doing so, and to demonstrate the capabilities of this technique, we determine the orbital periods of the CBPs in the above \kepler systems using the mechanism introduced in Kostov et al. (2016), compare the calculated periods with the corresponding true periods (obtained from photodynamical models of transits that occurred over several conjunctions), and examine the crucial effect of orbital precession. We also estimate the expected yield when the occurrence of multiple transits in a single conjunction is used to discover new transiting CBP candidates in the \tess\ data.

This paper is organized as follows. In \S 2, we present a brief review of the calculation of the orbital period of a CBP candidate using multiple transits in a single conjunction. In \S 3, we calculate the orbital periods of CBPs in the above-mentioned \kepler systems using this approach. Section 4 discusses the expected yield of detecting CBPs in the \tess\ data, and \S 5 presents a brief summary of the methodology for detecting \tess\ CBPs. Section 6 concludes our study by reviewing the results and discussing their scientific implications.

\section{Calculating the Orbital Period of Candidate CBPs}

In this section, we review the calculations of the orbital period of a CBP candidate detected in an eclipsing binary system that exhibits multiple transits in one conjunction. We note that these calculations are only possible for eclipsing double-lined spectroscopic binary stars because we need to know the locations of the stars with respect to the center of mass at any given time. For more details, we refer the reader to section 4 of Kostov et al. (2016) and to Schneider \& Chevreton (1990).

Briefly, using the measured transit times between two transits, ${\Delta t = |t_2 - t_1|}$, and the distance traveled by the CBP during this time interval, ${\Delta x = |x_2 - x_1|}$, the instantaneous, sky-projected $x$-component of the orbital velocity of the planet can be calculated from ${(V_{\rm calc})_x} = \Delta x/\Delta t$. Here, ${x_1}$ and ${x_2}$ are the sky-projected $x$-coordinates of the star being transited\footnote{The sky is in the $x-y$ plane, the observer is at $+z$, and the coordinate frame is barycentric. For details see chapter 2 of Hilditch (2001).}, and are derived spectroscopically from the binary's radial velocities at times ${t_1}$ and ${t_2}$ [for details, see Figure~\ref{fig:k34_LC} and also equations 3, 4 and 5 of Kostov et al. (2016)]. Using this velocity and the total mass of the binary, ${M_{\rm bin}}$ (also derived from binary's radial velocities), and assuming co-planarity between the CBP and the host binary, the orbital period of the CBP $({P_{\rm calc}})$ can be obtained from

\begin{equation}
{P_{\rm calc} = - \frac{2 \pi G M_{\rm bin}}{{(V_{\rm calc})}_{x}^3}  
\left[\frac{e \sin\omega + \sin(\theta + \omega)}{\sqrt{(1-{e^2})}}\right]^{3}}\,.
\label{eq:vel}
\end{equation}

\noindent 
In this equation, $e$, $\theta$, and $\omega$ are the eccentricity, true anomaly, and argument of pericenter of the CBP orbit; at conjunction, $\sin(\theta + \omega)\approx -1$. 

Kostov et al. (2016) applied this equation to the Kepler-1647 system and, assuming circular orbit, derived an orbital period of 1030 days for its CBP. This value is within ${\sim5\%}$ of the true orbital period of the planet, obtained from the comprehensive photodynamical modeling of this system. Such a level of accuracy was a remarkable achievement in using equation (1) as only less than 5 days of \kepler data had been utilized in the calculations. However, as explained in the next section, this impressive level of accuracy is mainly due to the fortuitous phase of the orbital precession of Kepler-1647 b such that \vcalcx\ is close to the average velocity of Kepler-1647 b during its orbit.
To obtain a more general understanding of the level of accuracy of this approximation, we applied Equation (1) to all \kepler systems that show multiple transits in one conjunction, ensuring that the CBP's orbital precession is properly taken into account. In the Section 3, we present and discuss these cases.

\subsection{Constraints on the orbital eccentricities of CBP candidates from stability analysis.}

As shown by equation (1), an estimate of the orbital eccentricity of a CBP candidate is required in order to calculate its orbital period. However, the eccentricity and the argument of pericenter are not known {\it a priori}. To some extent, this complication can be overcome using a stability analysis to constrain the range of the planet's orbital eccentricity. 

\begin{figure}
    \centering
    \includegraphics[width=\linewidth]{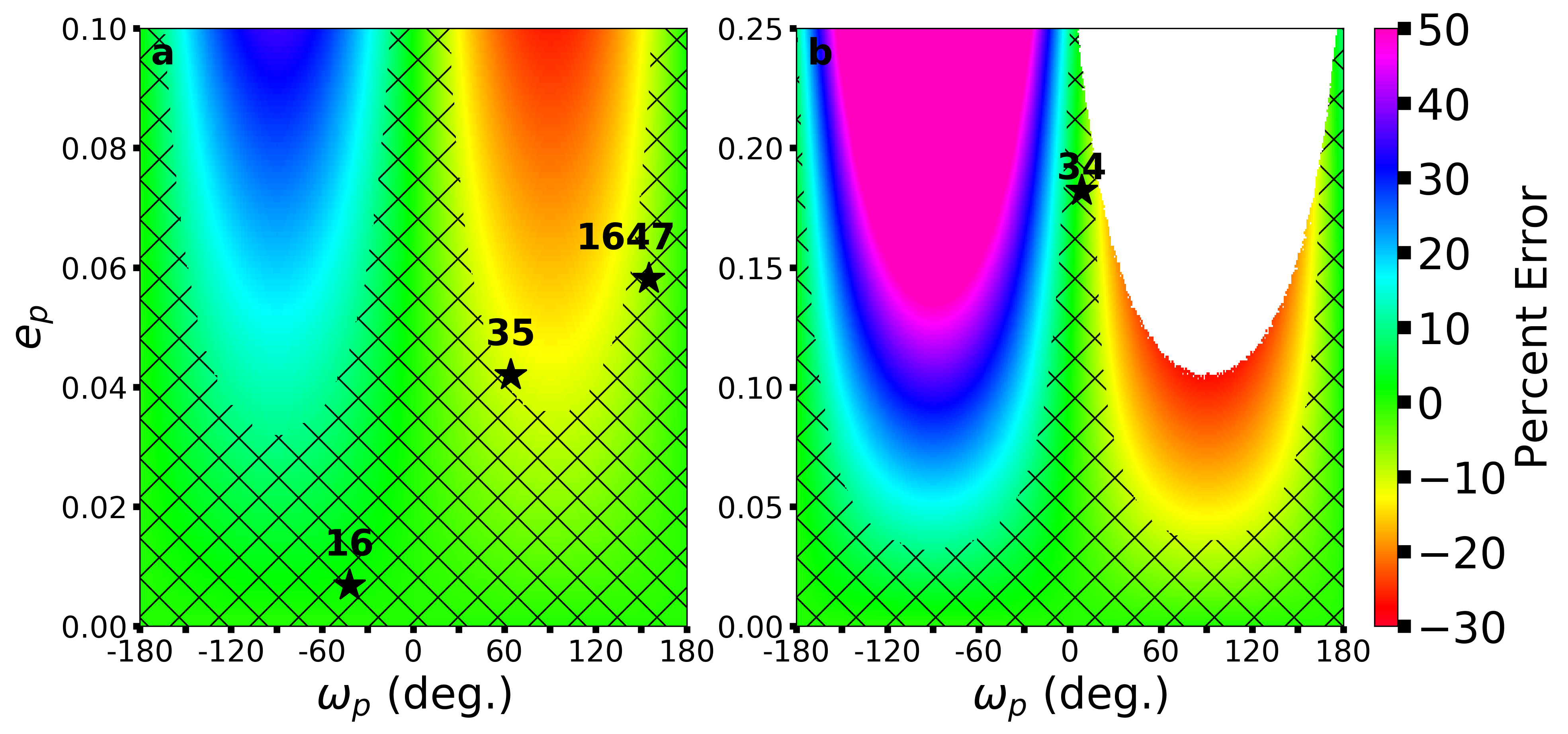}
    \caption{Percentage difference between circular-orbit periods and eccentric-orbit periods as calculated from Equation 1 (for a range of planetary eccentricity $(e_p)$ and argument of pericenter $(\omega_p)$). The star symbols correspond to Kepler-16b, -34b, -35b, and Kepler-1647b. Panel a) shows that most of the parameter space corresponds to less than $\sim10\%$ (hatched area) differences. Kepler-16b and Kepler-1647b result in differences that are substantially smaller than $\sim10\%$, while Kepler-35b is slightly higher ($\sim15\%$). Panel b) is identical to panel a), but extends to $e_p=0.25$ to accommodate for Kepler-34b. Here, about half of the parameter space corresponds to less than $\sim10\%$ differences and a significant portion can be excluded from an orbital stability analysis (white region).
    \label{fig:period_err}}
\end{figure}

This is demonstrated in Figure \ref{fig:period_err} where we compare the calculated orbital periods of Kepler-16b, -35b, -1647b, and -34b CBPs using Equation \ref{eq:vel} for circular and for eccentric orbits. As 12 of the currently known 13 transiting CBPs have orbital eccentricities smaller than 0.1\footnote{And 8 of these 12 have an orbital eccentricity smaller than 0.05.}, this value was chosen as the maximum range of the eccentricity we explored for Kepler-16b, -35b, and -1647b (panel a). Because the host binary for Kepler-34b is more massive and more eccentric compared to the other three systems, and so is the orbit of the CBP itself $(e= 0.18)$, we explored higher planetary eccentricities in panel b. We note that the forced eccentricities are low as well (see e.g., Eqn (38) of Leung \& Lee 2012).

For each combination of $(e,\omega)$ the orbital stability of the system was examined using the stability criteria and interpolation method developed by Quarles et al. (2018). All unstable orbits were removed (they correspond to regions shown in white in the figure). The hatched regions in each panel denote the $(e,\omega)$ range where the eccentric-orbit period differs from the circular-orbit period by less than 10\%---which is most of the parameter space for Kepler-16b, Kepler-35 b, and Kepler-1647 b systems. A significant portion of the parameter space for Kepler-34b can be excluded based upon orbital stability (Quarles et al. 2018), and the $(e,\omega)$ range spanning less than 10\% differences is substantial for this system as well.


This further strengthens the validity of our assumption for circular orbits when applying Equation (1). We note that this assumption is supported by the physics of planet migration. Specifically, given the low orbital eccentricities of the known transiting CBPs, these planets likely formed at large distances away from their host binaries and migrated to their current orbits through interactions with the circumbinary disk. Such interactions would inhibit the growth of the planets' eccentricities (e.g., Kley \& Haghighipour 2014). Thus based on both theoretical and observational arguments we believe that circular orbits are a safe assumption for CBPs.

\section{Application to {\it Kepler} CBPs}
\label{results}

Similar to calculating the orbital period of a transiting planetary candidate in a single-star system exhibiting a lone transit, in order to be able to use equation (1) to calculate the orbital period of a candidate CBP, simplifying assumptions such as circular orbits are needed. We note that 
the rich dynamical environments of close binary stars can have profound effects on the orbit of a CBP.
For instance, due to precession, the orbital elements of CBPs vary from one conjunction to the next and as a result, over one precession cycle, \vcalcx~will sample the full range of the planet's orbital velocity.

We note that a slight misalignment between the orbits of the CBP and the binary star can cause the former to not transit at every conjunction. Instead, in such cases, several consecutive transits may occur followed by long intervals where no transits will appear. To investigate the full effect of orbital precession on the variations in \vcalcx, we applied equation (1) to Kepler-like CBP systems where the orbits of the binary star and the planets were considered to be co-planar. In such systems, the CBP produces (at least) two transits at every conjunction. In order to examine the full amplitude of the precession-induced modulations in \vcalcx, we integrated these systems for one precession cycle of the CBP using the photodynamical model of Carter et al. (2011) and the parameters listed in their corresponding references. The model combines $N$-body simulations with appropriate light-curve models to simulate photometric time-series for direct comparison with the observations. Below we present the application of equation (1) to the four \kepler CBPs that produce multiple transits during one conjunction, assuming circular orbits and investing a full precession cycle.

\begin{figure}
\centering
\includegraphics[width=0.8\linewidth]{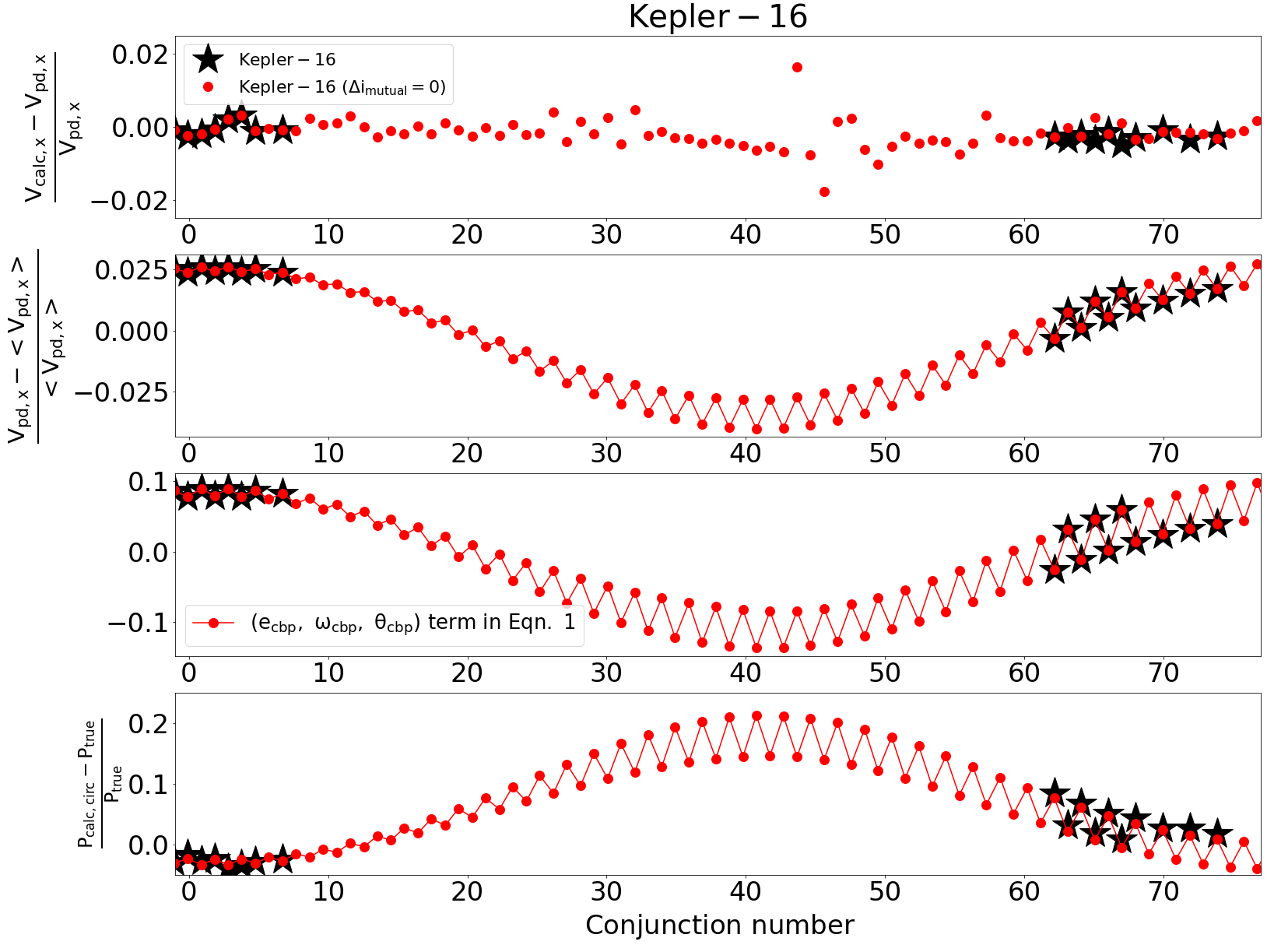}
\caption{{\it First panel from top}:  Relative comparison between the calculated orbital velocity of Kepler-16~b \vcalcx~and the instantaneous, photodynamically-derived velocity ${(V_{\rm pd})_x}$ at each conjunction producing two transits, and over a full orbital precession cycle. The red dots represent a fictitious co-planar Kepler-16 system and the black stars correspond to the actual Kepler-16 system (see Table \ref{tab:K16_times} for observed conjunction times). {\it Second panel}: Relative variation of the normalized velocity ${(V_{\rm pd})_x}$ over one precession cycle. {\it Third panel}: Same as second panel but for the ${\rm {[e \sin\omega + \sin(\theta + \omega))/(\sqrt{(1-{e^2})}]}^{3}}$ term in Eqn. 1; {\it Fourth panel}: Comparison between the calculated period of the planet ${\rm P_{calc, circ}}$ assuming circular orbit and the true period \ptrue.
\label{fig:k16}}
\end{figure}

Figure \ref{fig:k16} shows the results from the application of equation (1) to the Kepler-16 system. The orbit of the planet precesses around its binary with a period of ${\sim42}$ years, or approximately 67 planetary periods. During the time of the operation of {\it Kepler}, this CBP exhibited {\em{five}} cases of one-conjunction double-transit events, demonstrating that such observational signatures are not rare. As seen from the figure, \vcalcx~and ${(V_{\rm pd})_x}$ are within ${\sim0.5\%}$ of each other, demonstrating that not only is this method for calculating velocity from transit times and the traveled distance correct, but also remarkably precise. The second panel shows ${(V_{\rm pd})_x}$ normalized over the full precession cycle, indicating that the velocity varies by ${\sim5\%}$.  The third panel shows the normalized ${\rm {[e \sin\omega + \sin(\theta + \omega))/(\sqrt{(1-{e^2})}]}^{3}}$
term in Eqn. 1 over the full precession cycle.  The fourth panel shows a comparison between \pcalc~as calculated using equation (1),  assuming circular orbit, and the true period \ptrue  at each conjunction. On average, ${P_{\rm calc}/P_{\rm true}\approx1.1}$ over a full precession cycle and, overall, \pcalc~does not differ from \ptrue~by more than 20\%. 

The observed conjunctions, in terms of time of primary transit and time of secondary transit, and the calculated periods for Kepler-16, -34, -35, and -1647 are listed in Table \ref{tab:K16_times}. The corresponding variations in the calculated periods over a full precession cycle and assuming circular orbits are listed in Table \ref{tab:precession_variations}. 
Not unexpectedly, the systems with the larger eccentricities tend to fare worse when a circular orbit is assumed. Nevertheless, the relative error in the calculated period for the planet candidate ranges from an astonishingly good $\sim1\%$ to a respectable $\sim25\%$, which is still sufficient for population statistics.

\subsection{Lessons learned}

\begin{figure*}
\centering
\epsscale{0.75}
\plotone{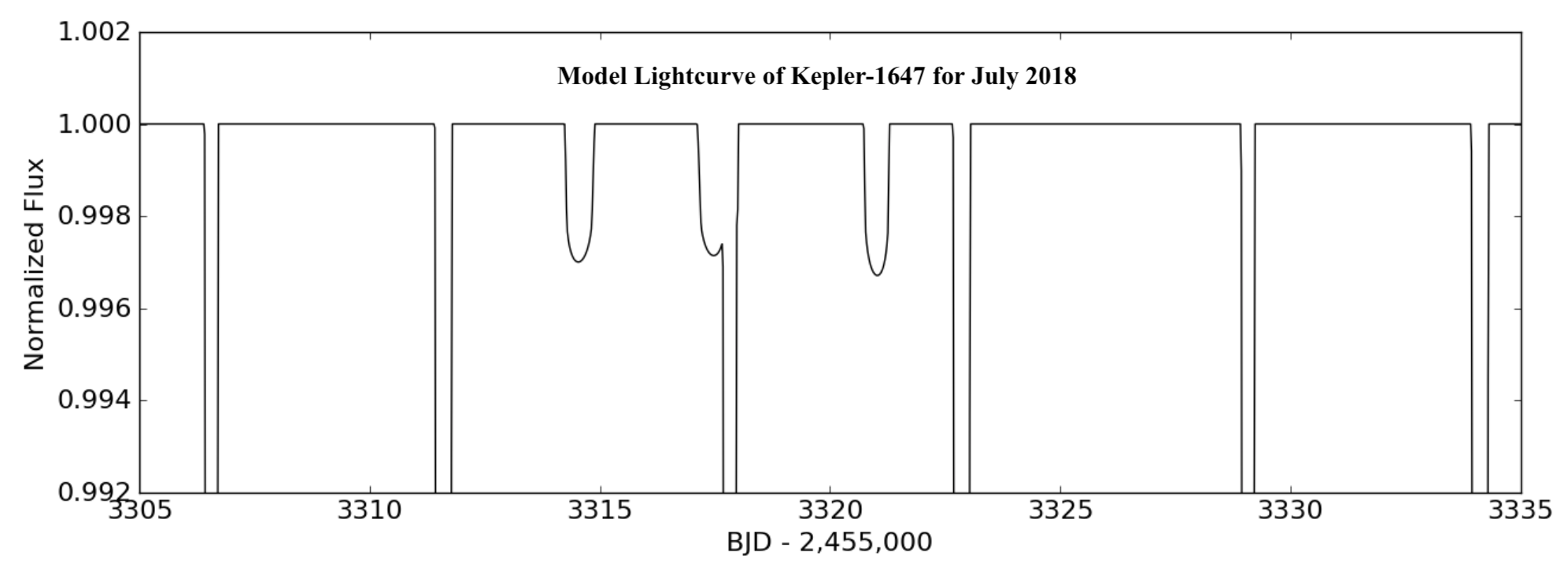}
\plotone{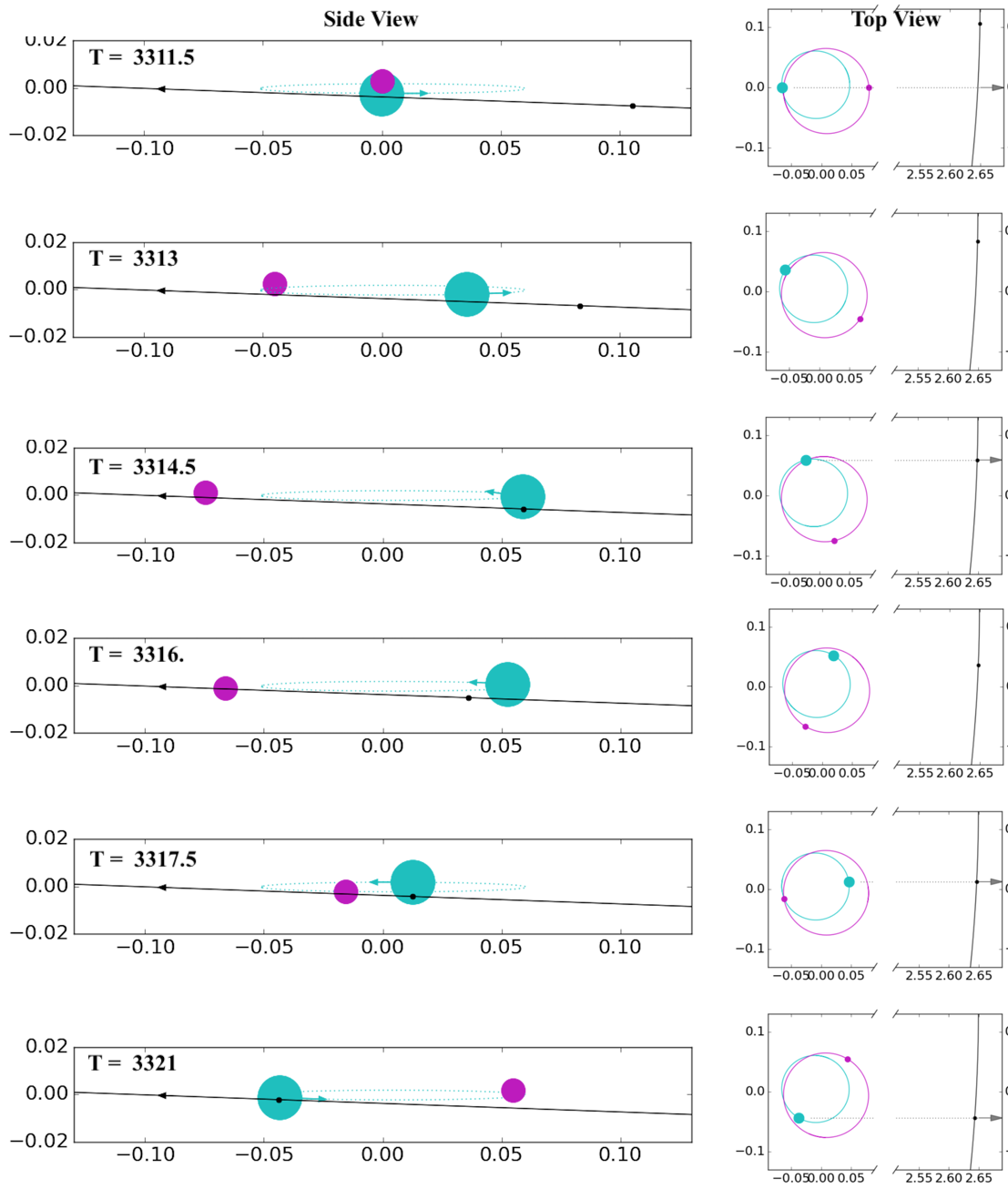}
\caption{{\it Upper panel}: Predicted light-curve for Kepler-1647 for July 2018, showing a triple 
transit of the planet across the primary star during a single conjunction. Such transits would easily fall within the ${\sim 28}$-day observing window of \tess\ and, given sufficient signal-to-noise, would be detectable. {\it Lower panels}: Orbital configuration for the above predicted triple transit. The sizes of the stars (magenta and cyan), their orbital configuration, and the orbit of the CBP (dotted line) are to scale. The size of the planet (black circle) has been exaggerated by a factor of 5 for viewing purposes. The observer is along the dashed gray line on the right panels. The units of the x- and y-axes of all lower panels are in AU.
\label{fig:1647_triple_transit}}
\end{figure*}

Kepler-1647 b exhibits yet another interesting feature of transiting CBPs---the occurrence of three (or more) transits during the same conjunction. This is demonstrated in Figure \ref{fig:1647_triple_transit} where we show the simulated light-curve of Kepler-1647~b for July 2018 when the planet must have transited the primary star three times during the same conjunction. In fact, such transits are common for Kepler-1647, and the CBP produces on average three transits across the primary and one across the secondary every third consecutive conjunction (in Figure \ref{fig:1647_triple_transit}, there is a transit across the secondary star, blended with the stellar eclipse near day 3318).

In general, in systems with circular and co-planar CBPs, two or more transits can occur across the same star during the same conjunction when:

\begin{equation}
{P_{\rm CBP} > P_{\rm bin}[(M_1/M_2) + 1]^3}
\label{eq:many_transits}
\end{equation}

\begin{figure*}
\centering
\epsscale{0.75}
\plotone{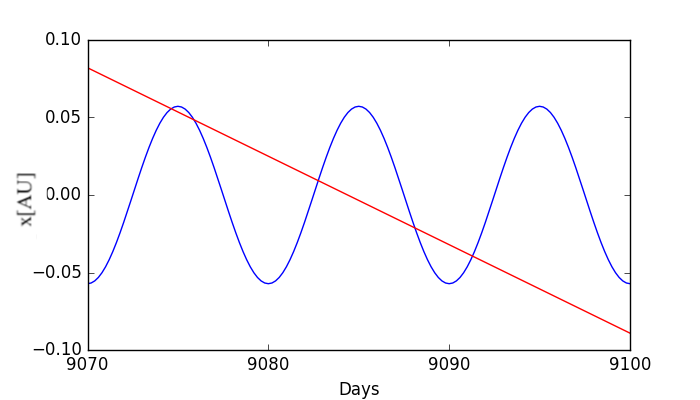}
\caption{An example quintuple transit of a CBP across the primary star in a simulated binary system composed of two solar-mass stars on a 10-day orbit and a CBP on a 20,000-day ($\sim$55-yr) orbit. The ordinate represents the sky-projected X-coordinate position of the primary star (in blue) and of the CBP (in red). The orbits of the binary and the CBP are aligned and the latter produces transits at every conjunction. The planet crosses the disk of the primary star five times during a single conjunction or stated more precisely, the primary star passes behind the planet five times, sometimes in a prograde and sometimes in a retrograde manner.
\label{fig:quintuple_transit}}
\end{figure*}

\noindent
Here, ${M_1}$, and ${M_2}$ are the masses of the primary and secondary stars, respectively, and ${P_{\rm bin}}$ is the period of the binary. Figure \ref{fig:quintuple_transit} shows a simulated example of such systems with quintuple transits across two Solar-mass stars in a 10-day binary with a CBP in a 20000-day orbit. In general, multiple transits across the same star can occur if the transverse velocity $V_x$ of the planet is slower than $V_x$ of the star (see also Fig. 2 in Deeg et al. (1997)).

One strong capability of equation (1) is that in such systems, any pair of transits can be used to calculate the orbital period of the CBP candidate. For instance, using the model light-curve of Kepler-1647 b shown in Figure \ref{fig:1647_triple_transit}, the calculated orbital period of the CBP is 1020 days for the transit pair T=3315 \& 3318, a calculated period of 1035 days for the transit pair T=3315 \& 3321, and 1050 days for transits T=3318 \& 3321, all within ${\sim5\%}$ of the CBP's true period.

It is important to note that the orbital eccentricities of Kepler-16, -35, and -1647 are close to zero, and during the Kepler observation of these systems, quantities $e$, $\omega$, and $\theta$ varied such that the term $e\, \sin \omega + \sin(\theta+\omega)$ in equation (1) stayed close to unity. In contrast, for the system of Kepler-34 both the host binary and the CBP have significant eccentricities (${e=0.52,\ e_{\rm CBP}\sim0.2}$). As a result, over one precession cycle the calculated \pcalc\ of the Kepler-34 CBP can be nearly a factor of 2 different from its true value (see Table \ref{tab:precession_variations}), thus pointing to the limitation of equation (1) when applied to such systems.

For completeness, we note that single-transit events observed from single stars can be used to estimate the orbital period of the planet as well. However, an expected uncertainty on the period of less than 10\% can be achieved for less than 10\% of such events expected from single TESS stars---even when assuming circular orbits (e.g.~Villanueva Jr. et al. 2019). For most of these events, the uncertainty will be much higher (see Fig. 4, Villanueva Jr. et al. (2019)), and can reach up to more than an order of magnitude. Eccentric orbits add another ${\sim50\%}$ of uncertainty. This is in contrast to the period uncertainties of transiting CBPs presented here, both for circular and for eccentric orbits.

\section{Expected CBP Yield from \tess}
\label{yield}

The currently known {\it Kepler} CBPs have a median orbital period of $\sim 175$ days (Socia et al. 2020), which is 6.3 times the \tess\ observing window of ${\sim 28}$ days (Ricker et al. 2015). With an observational duration only 1/6th the duration of an orbit, it is practically impossible to detect transiting CBPs in \tess\ all-sky data using the conventional method of observing of (at least) two conjunctions\footnote{Except near the ecliptic poles (${\sim 2.2\%}$ of the sky (Ricker et al. 2015)). However, the much small number of stars in that region will limit the number of CBP candidate detections.}. The occurrence of multiple transits in one conjunction allows us to circumvent this obstacle by applying equation (1) to the data obtained in a single observational window (corresponding to a single conjunction) to identify CBP candidates. In this section, we present the expected yield from using this method in the context of \tess\ observations.

To calculate the all-sky yield of detecting CBP candidates in \tess\ data, we assume that they will, statistically, present similar orbital and physical characteristics as the CBPs discovered by \kepler (i.e.\ Saturn-size planets with orbital periods of ${\sim100+}$~days). We also assume that the detection bias of \tess\ will be comparable to that of \kepler. Considering a 4-year \tess\ observing strategy (i.e., two years of primary mission plus two years of extended mission), and assuming a similar rate of detection of eclipsing binaries as by \kepler, Sullivan et al. (2015) have shown that \tess\ will observe about 476,000 eclipsing binaries (including hierarchical systems) brighter than $K_s = 15$ (their ``bright catalog''), with orbital periods between 0.5 days and 50 days. Using the analytic formalism of Li et al. (2016) and extrapolating from {\it Kepler}'s $\sim0.5\%$ CBP detection probability (12 CBPs out of ${\rm \sim2600}$ eclipsing binaries with orbital periods $<$50 days), we estimate that there will be $\sim2380$ potentially observable CBPs from \tess. 

\begin{figure}
    \centering
    \includegraphics[width=\linewidth]{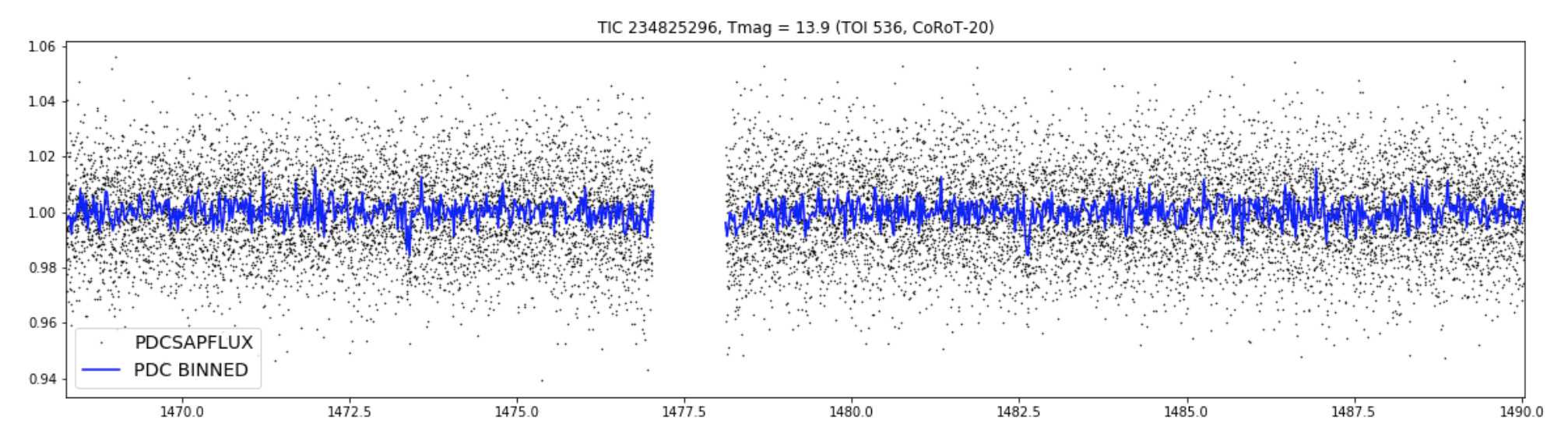}
    \caption{\tess\ light curve of TIC 234825296 (TOI 536, CoRoT-20), demonstrating that ${\sim1\%}$-deep transits are visible by eye for a target with ${T\approx14}$ mag. The dots show the 2-min cadence PDCSAP data, and the blue line represents the data binned to 30-min cadence. The depths of the two transits of TIC 234825296 are {\it smaller} than those of the transits of the CBP Kepler-16~b across its primary star.}
    \label{fig:Corot_20}
\end{figure}

We note that while most of these eclipsing binaries will likely be faint, i.e.~${\sim13-14}$ mag, \tess\ has already demonstrated its capability to detect individual transits of ${\sim1\%}$ depth for targets with ${T\approx14}$ mag. Such transits are also clearly visible even by eye, as shown on Figure \ref{fig:Corot_20} for the case of TIC 234825296 (TOI 536, CoRoT-20, ${T=13.9}$ mag), which is key for their detection for the technique discussed here. For context, the depths of both transits of TOI-536 are {\it smaller} than those of the CBP Kepler-16 b across its primary star (${\rm \approx14000\ ppm\ vs\ \approx17000\ ppm}$).

A few important points must be noted here. First, as this estimate is based on small numbers statistics, it may have large uncertainties. Second, the \kepler field of view was carefully selected to optimize exoplanet detection, which might have introduced a bias in the observed stellar population. Sullivan et al. (2015) note that, overall, the uncertainty of their EB estimates may be as large as ${\rm 80\%}$, especially at low galactic latitudes. We, therefore, adopt this uncertainty throughout the CBP yield presented here.

Scaling down to the duration of the all-sky \tess\ data (${\sim 28}$ days), the probability to detect at least one transit will be reduced by a factor of $6.3$ (the ratio of the CBP median orbital period to the observational window)\footnote{Sections with coverage longer than ${\sim 28}$ days will have higher detection probability.}. We therefore estimate that the Full Frame Image (FFI), all-sky, \tess\ data will yield $\sim380$ new CBPs. Assuming the same rate as in \kepler, 4 out of 11 of these CBP systems will produce two or more transits during one conjunction. This corresponds to $\sim140\pm110$ \tess\ CBPs expected to exhibit multiple transits during one conjunction and therefore suited for the analysis presented here. Numerical simulations are consistent with this estimate and confirm that the decrease in observed time does not correspond to a decrease in detection probability due to orbital precession. As a result, by analyzing \tess\ eclipsing binaries that exhibit multiple transits we will be able to increase the number of transiting CBPs by an order of magnitude. 

For the sake of completeness, we also present an estimate of the expected CBP yield from the two continuous viewing zones of \tess\ ($\sim 356$ days of observations). The fraction of the sky observed in these zones is $\sim 2.2\%$ (Ricker et al. 2015). Scaling down from the ${\sim476,000}$ all-sky EBs, these two regions contain $\sim10,000$ EBs, i.e. about four times as many as observed by the \kepler telescope. This will yield about $40\pm32$ CBPs. The first discovered TESS CBP in the continuous viewing zone, TOI-1338, has recently been reported by Kostov et al. (2020).

\section{Detection, Validation and Confirmation of \tess\ CBPs}
\label{discuss}

We expect the discovery of \tess\ CBP candidates to proceed as follows. First, visual\footnote{Potentially also automated searches such as Windemuth et al. (2019) and Martin \& Fabrycky (in prep)} inspections of the light-curves of \tess\ EBs will flag potential transit-like events. While this may sound like a daunting task, we note that the total data volume of \tess\ EBs is in fact comparable to the total data volume of \kepler EBs. That is, ~${\sim 476,000}$ \tess\ EBs ${\times \sim28}$ days of observations is only about a factor of 4 larger than ${\sim 2,000}$ \kepler\ EBs ${\times \sim1500}$ days of observations. A number of teams have inspected the latter many times over (e.g.~Welsh et al. 2012, Kostov et al. 2013, Armstrong et al. 2014, Martin \& Triaud 2014). 

If multiple, closely-spaced transits are detected in the light-curve of a \tess\ EB, the data will first be scrutinized for instrumental artifacts and/or astrophysical false positives due to, for instance, background EBs. Specifically, we will analyze the photocenter motion of the \tess\ images during the detected transits, and evaluate the probability for a background contamination following the methods of Kostov et al. (2020, Section 2.3). Next, the measured transit depths will be used to distinguish between a circumbinary third star, and a circumbinary planet. Passing these tests, a CBP candidate is identified. Next, assuming that the orbit of this candidate is circular and co-planar, equation (1) is used to estimate its orbital period. Photometric and spectroscopic observations will be necessary to obtain additional stellar eclipses (to examine eclipse timing variations), calculate the parameters of the binary star, and further rule out an eclipsing triple-star system (e.g.~through rigorous analysis for long-term trends in the measured radial velocity, or large amplitude eclipse timing variations). Note that archival photometric light curve data will be particularly valuable because \tess\ stars are generally bright, the eclipses of the stars are generally deep and easily detected, and because these data can immediately provide a substantial temporal baseline that is so helpful for determining the orbital period (and period changes) of the binary stars. A photodynamical analysis will be used to predict the times of future CBP transits. Further follow-up of these, if successful, will confirm the planetary nature of the CBP candidate. Finally, the \tess\ data, combined with the follow-up observations, incorporated into a full photodynamical analysis, will provide a comprehensive description of the size and orbital elements of the CBP. 

We note that some CBP candidates might be massive enough to perturb the binary and if the binary is bright enough to allow for RV follow-up, it might be easier to confirm these candidates through RVs instead of using photometric follow-up.
We note that there is also an ongoing survey to detect radial velocities of CBPs (BEBOP, Martin et al. 2019).

\subsection{Limitations.}

Using eclipsing binaries with multiple transits in one conjunction to search for CBPs presents the only known technique for identifying a large number of transiting CBP candidates in 28 days of \tess\ observations, and for estimating their orbital periods without an extensive follow-up effort. It is, however, necessary to mention that similar to any detection technique, this method is also based on certain assumptions that may introduce various difficulties and/or uncertainties. Specifically, these assumptions are as follows:

1) We assume the Kepler sample of CBPs is representative of the true sample. We also assume the masses, sizes and orbits of the host EBs are known. While this does take effort, it is based on established procedures and is routinely done with small telescopes and modest spectrographs. 

2) We assume that the ${\rm sin(\theta+\omega)}$ term in Equation 1 is -1 at conjunction. However, it will not be exactly -1 but will depend on the orbital phase of the star being transited.

3) We assume that Eqn. 1 is not affected by uncertainties in the binary's radial velocity measurements. However, these will affect the calculated mass of the binary $M_{bin}$ and thus \vcalcx\ such that, for example, 5\% uncertainty in \vcalcx\ will add 15\% uncertainty in \pcalc.

4) We assume future transits will occur near integer multiples of $P_{\rm calc}$ from the \tess\ transit, i.e. the ``moving target'' effect\footnote{The TTVs induced by the changing geometric configuration of the binary stars; see e.g.~Welsh \& Orosz 2018.} is neglected. This will propagate the uncertainty from point (2) into the prediction of the times of future transits. To obtain a more precise prediction, Monte Carlo photodynamical models with parameters spanning the uncertainties can (and will be) carried out to give a distribution for the predicted times.



\section{Conclusions}
\label{end}

Building upon the results of Kostov et al. (2016), we presented a study on utilizing multiple transits that occur during one conjunction to rapidly identify transiting CBP candidates in eclipsing binary systems. The strength of this analysis is in identifying the most likely CBP candidates and estimating an orbital period that is a reasonable approximation of its true period. The period estimate can be used as a guide for follow-up observations and for contributing to the population statistics of CBPs. As the effect can occur for an arbitrary eclipsing binary and CBP period, this approach is well-suited to identify \tess\ CBPs from both the all-sky data and the continuous viewing zones: long-period planets are not heavily disfavored. 

Capitalizing on the hundreds of thousands of eclipsing binaries that will be monitored by \tess\, we expect to be able to identify $\sim140\pm110$ transiting CBP candidates in the Full-Frame Images, and an additional $\sim40\pm32$ CBPs in the continuous viewing zones of \tess\ --- the first of which was just recently announced (Kostov et al. 2020). Thus multiple transits in one conjunction are ideal for obtaining the critical information---planet radius and estimated orbital period for a large number of circumbinary planet candidates from \tess\---necessary for follow-up observations and subsequent confirmation.


These discoveries will enable new CBP science through statistical studies of their occurrence rates, origin, and habitability, and will shed light on the formation and evolution of these objects. We will be able to characterize the population of CBPs in terms of i) the distribution of their orbital periods, sizes, (potentially) masses, eccentricities, and orbital inclinations; and ii) the distribution of the orbital periods, eccentricities, masses, metallicities and ages of their host binary stars. Furthermore, such a large sample of CBPs will allow detailed investigations of the correlations between their orbital and physical parameters, as well as comparisons between the populations of binary-star planets and single-star planets. Interestingly, this new approach can also probe planetary periods longer than even the year-long observations of continuous viewing zone of \tess. This will enhance the science return of the mission and maximize the long-period, temperate planet yield. Overall, \tess\ is the best instrument capable of detecting a large number of transiting CBPs in the coming years --- the only other comparable mission will not be launched before 2026 (PLATO, Rauer et al. 2013).

\acknowledgments

 We thank the referee for helping us improve this manuscript. NH acknowledges support from NASA XRP through grant number 80NSSC18K0519.  WFW and JAO gratefully acknowledge support from the NSF - this material is based upon work supported by the National Science Foundation under Grant No. (AST-1617004). We would like to thank John Hood, Jr. for his generous donation supporting this ``CBP 1-2 punch'' investigation, and exoplanet research at SDSU, in general.


\vspace*{0.5cm}

{\noindent Armstrong, D. J., Osborn, H. P., Brown, D. J. A., et al. 2014, MNRAS, 444, 1873 \\}
Bromley. B. C. \& Kenyon, S. J. 2015, ApJ, 806, 98 \\
Carter, J. A., Fabrycky, D. C., Ragozzine. D. et al. 2011, Science, 331, 562 \\
Deeg, H.-J., Martin, E., Schneider, J. et al. 1997, A\&AT, 13, 233 \\
Doyle, L. R., Carter, J. A., Fabrycky, D. C., et al. 2011, Science, 333, 1602 \\
Haghighipour, N. \& Kaltenegger, L. 2013, \apj, 777, 166 \\
Kley, W. \& Haghighipour, N. 2014, \aap, 564, 72 \\
Kley, W. \& Haghighipour, N. 2015, \aap, 581, 20 \\
Kostov, V. B., McCullough, P. R., Carter, J. et al. 2014, \apj, 784, 14 \\
Kostov, V. B., Orosz, J. A. Welsh, W. F. et al. 2016, \apj, 827, 86 \\
Kostov, V. B., Orosz, J. A., Feinstein, A. et al. 2020, in press \\
Li, G., Holman, M. J. \& Tao, M. 2016, \apj, 831, 96 \\
Martin, D. \& Triaud, A. 2014, MNRAS, 570, 91 \\
Martin, D. 2017, MNRAS, 465, 3235 \\
Pierens, A. \& Nelson, R. 2013, \aap, 556, 134 \\
Quarles, B., Satyal, S., Kostov, V. et al. 2018, \apj, 856, 150 \\
Rauer, H., Catala, C., Aerts. C. et al. 2013, arXiv:1310.0696v2 \\
Ricker, G. R., Winn, J. N., Vanderspek, R., et al. 2015, Journal of Astronomical Telescopes, Instruments, and Systems, 1, 014003 \\
Schneider, J., 1994, P\&SS, 42, 539 \\
Socia, Q. J., Welsh, W. F., Orosz, J. A. et al. 2020, 159, 94 \\
Sullivan, P. W., Winn, J. N., Berta-Thompson, Z. K. et al. 2015, \apj, 809, 77 \\
Sutherland, A. P. \& Fabrycky, D. C. 2016, \apj, 818, 6 \\
Villanueva, S. et al. 2019, \aj, 157, 84 \\
Welsh. W. F., Orosz, J. A., Carter, J. A. et al. 2012, Nature, 481, 475 \\
Welsh. W. F., Orosz, J. A., Short, D. R. et al. 2015, \apj, 809, 26 \\
Welsh, W. F. \& Orosz, J. A. 2018, {\it Handbook of Exoplanets}, Edited by Hans J. Deeg and Juan Antonio Belmonte. Springer Living Reference Work, ISBN: 978-3-319-30648-3, 2017 \\
Windemuth, D., Agol, E., Carter, J. et al. 2019, MNRAS, 490, 1313

\clearpage

\clearpage

\clearpage

\clearpage

\clearpage

\clearpage

\clearpage
\begin{table}[t]
\centering
\caption{Observed conjunctions and corresponding calculated period for Kepler-16, -34, -35, and -1657 assuming circular orbits.} \label{tab:K16_times}
\begin{tabular}{cccc}%
\hline
\hline
Conjunction & Primary Transit & Secondary Transit & Calculated Period \\
 & [BJD-2,455,000] & [BJD-2,455,000] & [days] \\
\hline
\multicolumn{4}{c}{\em Kepler-16~b (\ptrue$\approx229$ days)} \\
1 & -26.57926 & -18.40544 & 229.9 \\
2 & 203.7 & 195.3 & 228.3 \\
3 & 425.2 & 433.4 & 228.5 \\
4 & 655.46 & 647.12 & 225.5 \\
5 & 876.98 & 885.2 & 226.2 \\
\hline
\multicolumn{4}{c}{\em Kepler-34~b (\ptrue$\approx289$ days)} \\
1${^\dagger}$ & -- & -- & -- \\
2 & 508.3314 & 513.889 & 247.42\\
3 & 790.0313 & 797.4079 & 234.34\\
4 & 1076.446 & 1079.716 & 225.85\\
5 & 1366.827 & 1362.168 & 213.95\\
\hline
\multicolumn{4}{c}{\em Kepler-35~b (\ptrue$\approx131$ days)} \\
5 & 586.44 & 582.0 & 142.1\\
6 & 712.32 & 710.4 & 142.2 \\
\hline
\multicolumn{4}{c}{\em Kepler-1647~b (\ptrue$\approx1108$ days)} \\
1${^\ddagger}$ & -- & -- & -- \\
2 & 1104.952& 1109.264 & 1027.6 \\
\hline
\hline
\multicolumn{4}{l}{${\dagger}$ Secondary transit too shallow to be detected.} \\
\multicolumn{4}{l}{${\ddagger}$ Primary transit in data gap.}
\end{tabular}
\end{table}

\clearpage
\begin{table}[t]
\centering
\caption{Variations in the calculated orbital periods of Kepler-16, -34, -35, and -1657 (assuming circular orbits) over a full precession cycle.} \label{tab:precession_variations}
\begin{tabular}{c|cccc}%
\hline
\hline
CBP & Precession period & Precession period &  & \\

& [Years] & [CBP orbital periods] & 
${\rm (\frac{P_{calc,circ}}{P_{true}})_{max}}$ & 
${\rm (\frac{P_{calc,circ}}{P_{true}})_{min}}$ \\
\\
\hline
{\em Kepler-16~b} & $\approx42$ & $\approx67$ & $\approx1.2$ & $\approx1.0$  \\
{\em Kepler-34~b} & $\approx65$ & $\approx80$ & $\approx1.7$ & $\approx0.5$  \\
{\em Kepler-35~b} & $\approx21$ & $\approx60$ & $\approx1.1$ & $\approx0.8$  \\
{\em Kepler-1647~b} & $\approx7000$ & $\approx2300$ & $\approx1.2$ & $\approx0.8$ \\
\hline
\end{tabular}
\end{table}

\end{document}